\documentclass[sigconf,10pt]{acmart}

\usepackage[english]{babel}
\usepackage{blindtext}
\usepackage{amsfonts}
\usepackage{color}
\usepackage{graphics}
\usepackage{graphicx}
\usepackage{url}
\usepackage{subcaption}
\usepackage{xspace}
\usepackage{tabularray}

\newcommand{\ricsys}{RICworld\xspace}
\newcommand{\twin}{DigitalTwin\xspace}
\newcommand{\edgeric}{EdgeRIC\xspace}

\renewcommand\footnotetextcopyrightpermission[1]{} 
\setcopyright{none}

\acmDOI{}

\acmISBN{}

\acmPrice{}

\begin{document}
\title[EdgeRIC: Realtime Optimization and Control in NextG Networks]{EdgeRIC: Empowering Realtime Intelligent Optimization and Control in NextG Networks}
%
%
%
%
\author{Woo-Hyun Ko$^{1}$, Ushasi Ghosh$^{2}$, Ujwal Dinesha$^{1}$,  Raini Wu$^{2}$,\\ Srinivas Shakkottai$^{1}$ and Dinesh Bharadia$^{2}$   \\
\normalsize{{$^{1}$ Texas A\&M University, TX, USA, $^{2}$ UC San Diego, CA, USA}}\\
{\{whko,\ ujwald36,\ sshakkot\}@tamu.edu \quad \{ughosh,\ rainiwu,\ dineshb\}@ucsd.edu}}


\renewcommand{\shortauthors}{Woo-Hyun Ko et al.}

\begin{abstract}
Radio Access Networks (RAN) are increasingly softwarized and accessible via data-collection and control interfaces.   RAN intelligent control (RIC) is an approach to manage these interfaces at different timescales.  In this paper, we develop a RIC platform called \ricsys, consisting of (i) \edgeric, which is colocated, but decoupled from the RAN stack, and can access RAN and application-level information to execute AI-optimized and other policies in realtime (sub-millisecond) and (ii) \twin, a full-stack, trace-driven emulator for training AI-based policies offline.   We demonstrate that realtime \edgeric\ operates as if embedded within the RAN stack and significantly outperforms a cloud-based near-realtime RIC (> 15 ms latency) in terms of attained throughput.  We train AI-based polices on \twin, execute them on \edgeric, and show that these policies are robust to channel dynamics, and outperform queueing-model based policies by 5\% to 25\% on throughput and application-level benchmarks in a variety of mobile environments.

\end{abstract}

\maketitle

\section{Introduction}

NextG cellular networks must support applications ranging from interactive streaming media, AR/VR, control of robot systems, to industrial IoT 4.0.  Many of these applications depend on tightly coupled chains of sensing, communication, and computation that requires wireless communication with low-latency, high-throughput, and high-reliability guarantees hitherto unavailable.  The requirements are sometimes hard to quantify without contextual information about applications, such as nearness to stalling (media streaming), viewing angle (AR/VR), pose and location (robot control), age of information (industrial IoT).  The problem is compounded by the fact that wireless links are fast-changing across diverse spectrum bands and user mobility patterns.  To meet these requirements, traditional notions of tightly segregated layered networking are being revised in favor of highly adaptable cross-layer optimization of wireless resources.

\begin{figure}[htbp]
\vspace{-0.1in}
\begin{center}
\includegraphics[width=0.6\linewidth]{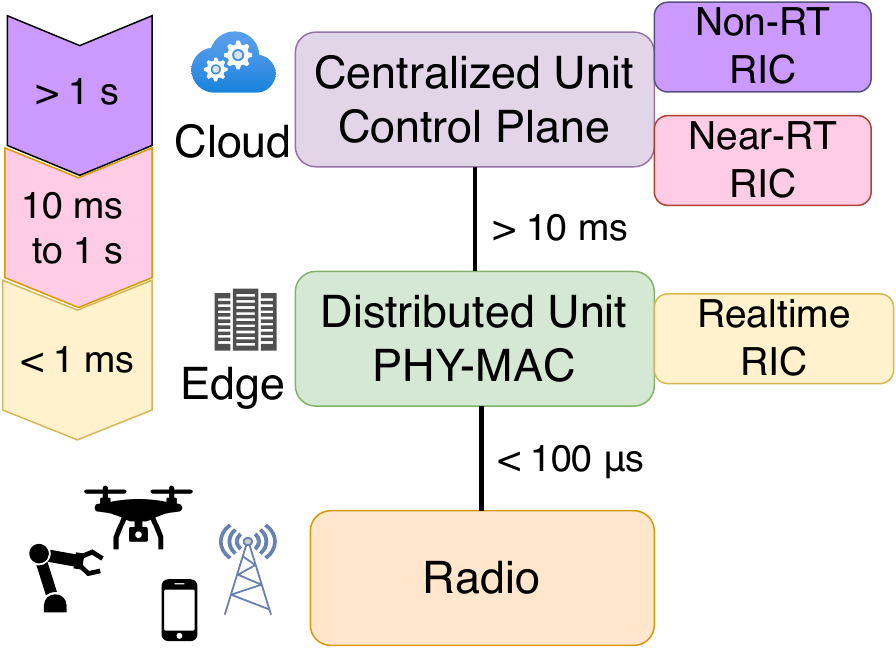}
\caption{Timescales of RAN Intelligent Control for O-RAN. We desire realtime control at a latency < 1 ms.}
\label{fig:timescales} 
\end{center}
\vspace{-0.1in}
\end{figure}

NextG networks have the potential for autonomous adaptation to achieve such cross-layer optimization.  With the advent of O-RAN (Open Radio Access Network), NextG cellular networking is being powered by softwarization and disaggregation at all layers, enabling both the ability to run the RAN stack on a variety of distributed compute, and the means to monitor and control it via newly established interfaces as shown in Figure~\ref{fig:timescales}.  In addition, the notion of RAN Intelligent Control (RIC)  has arisen in parallel as an approach to managing these new interfaces. RICs are decoupled from the time-sensitive RAN stack, and the hope is for them to access both application layer and RAN-level information to provide cross-layer decision and control, as well as AI/ML capabilities to the RAN stack for supporting diverse and demanding applications.

The RAN-stack operates in realtime at a transmission time interval (TTI) of 1 ms or less (based on the 5G standards)  i.e., it has to take control decisions at each ms.  Recent attempts at RIC have been cloud-compute-based (i) near-realtime RIC (near-RT RIC), supporting a feedback loop to RAN stack at a timescale of 10 ms to 1 s, and (ii) non-realtime RIC (non-RT RIC), supporting a feedback loop at a timescale > 1 s. This cautious approach is taken to avoid impacting the time-sensitive tasks that must be completed within each TTI in the RAN stack.  Breaching physical layer and medium access control (PHY-MAC) task-latency guarantees can lead to deleterious behaviors such as user equipment (UE) detachment.   Hence, there is a mismatch between the realtime operations of the RAN and the current near-RT or non-RT RIC solutions. 

The mismatch in RAN-level events and near-RT and non-RT RIC timescales implies that they can only undertake cross-layer decisions for prioritization of specific applications at a coarse level via resource slicing at the RAN. For instance, near-RT RIC may assign a streaming application to a resource-rich RAN slice, which might have the desired effect of reducing stalls because the RAN-level scheduler (which is oblivious to the application) provides low latency or high throughput using more resources available in that slice.  However, such coarse slicing approaches could quickly  become untenable with increasing loads being placed on wireless access networks by demanding applications, all of which cannot be assigned to such resource-rich slices.  

Providing tight performance guarantees to demanding applications while maintaining resource efficiency implies that decision-making must be at the TTI timescale of underlying channel variations. Thus, truly transformative solutions for algorithmic control of PHY-MAC will require a \emph{realtime (TTI-scale)} RIC located on edge compute near the radio head.

\subsection*{Main Contributions}\looseness-1

In this paper, we propose \ricsys, which can train and deploy policies for cross-layer information access and control at the sub-millisecond timescale.  \ricsys consists of two major  components, namely, (i) a monitoring and control microservice called \edgeric, and (ii) an emulator called \twin.  \edgeric is a microservice on edge compute, which is decoupled from the RAN stack and connects with it over an O-RAN like standard to execute AI-optimized and other policies in realtime (sub-millisecond). \twin\ is a cloud-compute based emulator, which enables the design, training and deployment of AI-optimized algorithms via full-stack emulation with multiple users and applications.  \ricsys can also leverage existing near-RT and non-RT RIC located on cloud-compute via an in-built information exchange.

\vspace{0.05in}\noindent\textbf{Design Decisions}\vspace{0.05in}\\
\emph{Where should \edgeric\ reside?}  An AI-optimized policy embedded within the RAN can access information in realtime, but also has a chance of breaking the latency-sensitive RAN stack, requires an application-aware RAN, and would require re-compiling the RAN for each algorithm update.  On the other hand, decoupling completely from the RAN via cloud-based near-RT RIC has the issue of delayed information from the RAN over jittery, non-RT network links.   Our key idea is that by co-locating with, but decoupling from the RAN via a realtime information exchange,  \edgeric cannot impinge on latency-constrained RAN tasks, while yet accessing information and providing control in realtime.  \edgeric hosts realtime applications called $\mu$Apps with control policies that are agnostic to the RAN implementation.  The design is fault tolerant in that an absence of input from \edgeric means that the RAN continues with its usual operation.

\emph{What should be \edgeric 's action space?}   Many queueing systems have highly structured optimal scheduling policies that take the form of deciding the relative weight of one queue vs. another---a notion called indexing, with the Whittle index being the most well-known such policy class.  In a likely first attempt in cellular networks, we leverage this observation to utilize a weight-based abstraction of control, under which each UE is assigned a weight by \edgeric at each TTI, which the RAN stack uses to determine their relative importance to perform resource allocation decisions.   We then provide OpenAIGym compatibility for $\mu$Apps to execute standardized AI-optimized policies that output these weights.

\emph{How do we train cross-layer AI-optimized algorithms?}   AI algorithm training in a production system can result in excessive compute loads and extended training times.  We equip \ricsys with \twin that resides on cloud-compute at non-RT RIC, under which the core, RAN, RICs and multiple UE modules and their applications can be run over a trace-based emulation of channel quality index (CQI).   AI/ML models are robustly trained under a variety of channel conditions, and are then sent for adaptation and realtime execution in $\mu$Apps on \edgeric.   Thus, $\mu$Apps are able to update their policies in runtime as training and adaptation continue.

\vspace{0.05in}\noindent\textbf{Performance Evaluation}\vspace{0.05in}\\
We evaluate \ricsys using the open source srsRAN stack by connecting to it via a custom variant of the O-RAN E2 application protocol (E2AP). We develop a custom scheduler placed within the srsRAN stack to perform proportional resource allocation using weights generated by \edgeric.   We show that the RAN to \edgeric round trip time (RTT) has a median of only about 100$\mu$s, and that scheduling algorithms operating at the TTI-timescale embedded within the RAN stack or as $\mu$Apps show near-identical spectral efficiency, i.e., \edgeric\ operates as if from within the RAN stack and is capable of realtime decision and control.

Next, we develop AI-optimized $\mu$Apps to provide resource allocation decisions each TTI.  The challenge is to complete measurement, policy execution and resource allocation within the TTI.   We train $\mu$Apps via domain-randomized reinforcement learning (RL) over \twin\ to optimize performance of two types of applications, namely (i) throughput maximization, (ii) streaming media with stall minimization over a variety of channels.   Training takes about  {ten minutes,} while execution takes {less than 150 $\mu$s}.  Hence, policy training and realtime feedback control are feasible on \ricsys.

Finally, we conduct over-the-air experiments and trace-based emulations with devices on a turntable, mobile robots, cars and drones.  Our baselines are well-established, queueing-model based algorithms, such as max-weight scheduling.  We first show that AI-optimized polices using the weight-based abstraction yield about 5\% throughput gain over classical scheduling.  We then show that in an environment where media streamers co-exist with other applications loading the system, \ricsys can provide over 25\% decrease in stalls.  We conclude that AI-based realtime cross-layer optimization of cellular network resources is both achievable and valuable.
\section{Related Work}

There has been significant recent work on developing near-RT RIC approaches, several of which use AI-optimized procedures for RAN control. For instance, Scope~\cite{bonati2021scope} provides a containerized approach for instantiating cellular network elements, an emulation module for testing in real-world scenarios, a data collection module for AI/ML applications, and APIs for control of network functionalities. The ColO-RAN \cite{coloran} platform is a publicly available Open RAN solution that is based on the world's largest network emulator, Colosseum \cite{colosseum}. \cite{nextGintelligence} conducted experiments on Colosseum to show the integration of a similar platform with deep RL agents. The focus of these works is on near-RT RIC based control, under which RAN resources are sliced in near-RT via an xApp and RAN-embedded schedulers are assigned to each one.

Other work explores the application space possible with near-RT RIC and non-RT RIC.  In this context, \cite{dhillonoranintelligence} proposes an intelligent radio resource management scheme which leverages LSTM neural networks, non-RT and near-RT compute platforms, to learn the spatial pattern of data traffic and predict possible congestion.  Based on these predictions, radio resources are dynamically re-allocated to prevent congestion and ensure optimal network performance.   \cite{oranE2}  presents a software-defined radio testbed featuring an open-source 5G system that interacts with the O-RAN near-RT RIC through standard interfaces with two xApps for RAN slicing.

Moving closer to the TTI timescale,  ChARM~\cite{baldesi2022charm} presents spectrum selection based on supervised learning over IQ samples utilizes data collection in realtime, but with control at near-RT RIC.  For even finer control, \cite{dapps} proposes an architecture for integrating distributed applications (dApps) into the O-RAN, and present simulations on the possible benefits of network intelligence at realtime (<10 ms).   Further, \cite{mobisys} provides a conceptual messaging approach for enabling realtime RIC (<1 ms) and presents a feasibility study.    In a similar framework, FlexRAN~\cite{flexran} provides a software-defined RAN platform, under which a master controller communicates with agents embedded in the RAN stack.  The architecture requires agents tailored to the LTE stack, rather than functionally isolating them as O-RAN standards suggest.  Although the master controller can host a MAC scheduler, FlexRAN does not show the ability to train or utilize AI-optimized policies while maintaining realtime constraints.   FlexRIC~\cite{flexric} is a more modular variant of FlexRAN.  The focus is on simplifying the 5G near-RT RIC architecture, using the same agent-controller approach of FlexRAN.  

In the applications domain, streaming media has received much attention for AI-optimized control.  For instance, AI/ML for choosing video streaming rate selection is considered in ~\cite{mao2017neural,huang2018qarc,zhang2019drl360,xiao2019deepvr, rate-utility} from the server's perspective, and assumes that the network provides packet delivery with certain statistics that can be learned.  \cite{bhattacharyya2021qflow} studies optimal policies
when the network can be controlled in the context of WiFi-based access.  
Here, non-RT reconfiguration of WiFi flow priorities using AI-optimized policies is shown to improve streaming performance.
\cite{ha2rs} investigates a system for streaming high-quality augmented reality (AR) content over wireless networks that experience rapid fluctuations in performance.\looseness-1

In contrast to the above work, ours is a simple, lightweight, decoupled approach towards ensuring that TTI-level synchronized policies can be trained in non-RT and executed in realtime.  Specifically, we show that our approach provides the ability to train robust cross-layer optimized policies in non-RT and a guarantee of completing the full feedback loop from sensing, AI-based policy execution and control within each TTI, and are the first to verify our claims while running full stack over-the-air experiments on mobile nodes.


\section{\ricsys  Concept Architecture}
\label{sec:concept}\looseness-1

We develop a concept architecture for \ricsys.   Our goals for its constituents, EdgeRIC and \twin\ are as follows:

\vspace{0.05in}\noindent \textbf{O-RAN Compatibility:}  EdgeRIC must be consistent with O-RAN messaging standards and \ricsys\ must provide compatibility with near-RT and non-RT RIC.  This will ensure ease of deployment and integration with existing RICs.

\vspace{0.05in}\noindent \textbf{Realtime:}  EdgeRIC must be able to access RAN state information and provide control actions at each TTI.  This will enable it to respond to a fast-changing wireless channel.

\vspace{0.05in}\noindent \textbf{Robustness:}  EdgeRIC must not have any blocking elements that could slow down the RAN stack.  This will ensure that any missing or delayed state or control does not break the tight constraints of TTI-scale RAN operations. 

\vspace{0.05in}\noindent \textbf{Reconfigurability:}  EdgeRIC must be able to update control polices in runtime.  This will permit adaptation in response to data collected without having to restart the RAN or RIC.

\vspace{0.05in}\noindent \textbf{Cross-layer Awareness:}  Apart from the RAN stack, EdgeRIC must be able to communicate with the applications running on UEs or other network elements.   This will enable cross-layer optimization while accounting for application state.

\vspace{0.05in}\noindent \textbf{Support for AI/ML Workflows:} EdgeRIC must support standardized AI/ML codebases that might require offline, online, simulation or emulation-based data collection and training.  This will ensure compatibility with a wide variety of AI/ML approaches for both inference and control. 

\vspace{0.05in}\noindent \textbf{Emulation Environment:}  \twin\ must operate over an identical stack of RAN, core, RIC(s), UEs and applications.  It must accept data digests from over-the-air execution to ensure a realistic wireless and application environment. Queueing-model based or AI-optimized policies should be directly transferable to EdgeRIC without modification.

\begin{figure}[!t]
\vspace{-0.1in}
\begin{center}
\includegraphics[width=\linewidth]{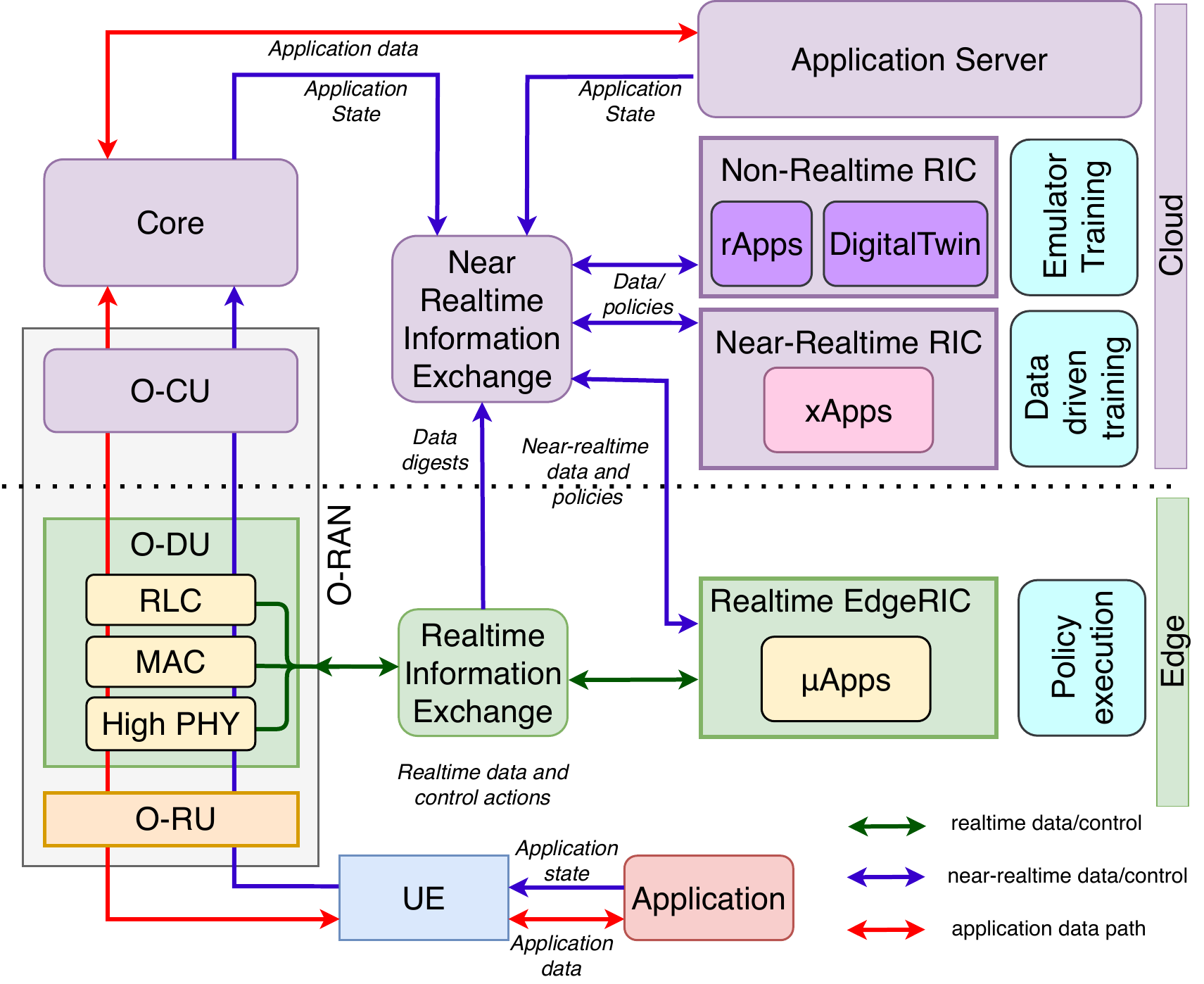}
\vspace{-0.2in}
\caption{\ricsys\ concept architecture, showing \edgeric, \twin\ and integration into O-RAN. 
}
\label{fig:RIC-6} 
\end{center}
\end{figure}

\subsection{Realtime O-RAN Connectivity} 
EdgeRIC is illustrated in Figure~\ref{fig:RIC-6}, where we have shown it within the O-RAN architecture.  O-RAN consists of a radio unit  along with  disaggregated microservices that perform the RAN functions.  These microservices are divided across edge compute (near the radio) and cloud compute resources, based on the required latency targets.    The components of the O-RAN stack are as follows:  (i) RF Frontend: Open Radio Unit (O-RU), (ii) Edge Compute: Realtime components at the Open Distributed Unit (O-DU) supporting High-PHY, MAC and radio link control, and  (iii) Cloud Compute: Open Centralized Unit (O-CU) with control and management functions.  The final element is (iv) Cloud Compute: 5G Core, supporting management, billing and Internet gateway functions.  

The O-RAN standard also provides specifications for two cloud-hosted microservices for (i) {near-RT RIC}, which supports xApps for policy adaptation at near-realtime (10 ms - 1 s), and (ii) {non-RT RIC}, which supports rApps providing microservices management and data analytics.  The standard provides protocols for the near-RT and the non-RT RIC to communicate with the O-RAN stack and with each other.    In particular, the E2 Application Protocol (E2AP) operates over SCTP and provides pub-sub and on-demand messaging between RAN and near-RT RIC at near-RT latency.

We align our architecture with O-RAN, and conceptualize EdgeRIC as a microservice positioned at the O-DU for physical proximity to realtime RAN functionalities at the PHY-MAC level.   EdgeRIC supports $\mu$Apps representing realtime policy execution  (TTI latency), and they can obtain RAN state information and apply control actions.  EdgeRIC connects to the O-RAN stack via a realtime information exchange, which we visualize as a realtime-E2 protocol (RT-E2), which operates over inter-process communication (IPC) between microservices to ensure realtime latencies.  

Our architecture also provides a near-RT information exchange, which can be used for data collection and sharing between RT, near-RT and non-RT RIC.  This data can include policies that are developed at one of the RICs and is to be updated or executed at one of the others.  The near-RT information broker can be visualized as a database that also provides the ability to store data digests for post processing.

\subsection{Decoupled Execution, Runtime Updates and Cross-layer Optimization}\looseness-1

The O-RAN stack must satisfy tight guarantees on PHY-MAC task completion times so as to maintain synchronization across all processes that have to be completed in each TTI.    Breaching PHY-MAC latency guarantees can lead to unstable performance or deleterious behaviors such as UEs detaching.   Although EdgeRIC operates in realtime, it is not integrated into the O-RAN PHY-MAC microservices at the O-DU.  This allows us to decouple EdgeRIC operations with those of the O-DU stack.  It can thus run on separate CPU cores from the O-DU microservices.  The realtime information exchange ensures that the O-RAN stack never is in a blocking state waiting for input from EdgeRIC.  Furthermore, events at EdgeRIC are driven by the TTI clock ticks from the O-DU, resulting in EdgeRIC being synchronized  to the events at the O-DU stack.

Our decision to instantiate EdgeRIC as a decoupled realtime microservice at the O-DU enables it to support a variety of $\mu$Apps that can be brought up and torn down at will.  Each such $\mu$App can utilize the realtime and near-RT information exchanges to obtain data to perform analytics, or to execute policies to generate control decisions.    Since the O-DU stack is non-blocking on inputs from EdgeRIC, these $\mu$Apps can be modified in runtime as data is collected and policies running on them need to be updated.

The decoupled $\mu$Apps-based architecture is also compatible with cross-layer information sharing with EdgeRIC.  While $\mu$Apps utilize RT-E2 messages for communicating with the O-DU, they can also use other protocols in near-RT or non-RT for obtaining information about other elements of the 
applications stack.  These protocols could include OpenFlow (network layer messaging), or a variety of application-level pub-sub protocols such as ROS (for robot sensing and control streams) or message queueing protocols (obtaining state information from AR/VR or media streaming applications).    Hence, the EdgeRIC architecture can easily incorporate cross-layer policies that optimally control the PHY-MAC stack.
 
\subsection{Emulator and AI/ML Workflow}



EdgeRIC directly enables support for optimization based approaches to modulation, coding and queuing that are designed around well studied, substantiated, and tractable models.  For instance, we can immediately instantiate approaches such as proportionally fair~\cite{sesia2011lte} or max-weight~\cite{tassiulas1990stability} scheduling across UEs on a per TTI basis with execution in RT, as if embedded within the RAN stack.

\begin{figure}[!t]
\vspace{-0.1in}
\begin{center}
\includegraphics[width=0.8\linewidth]{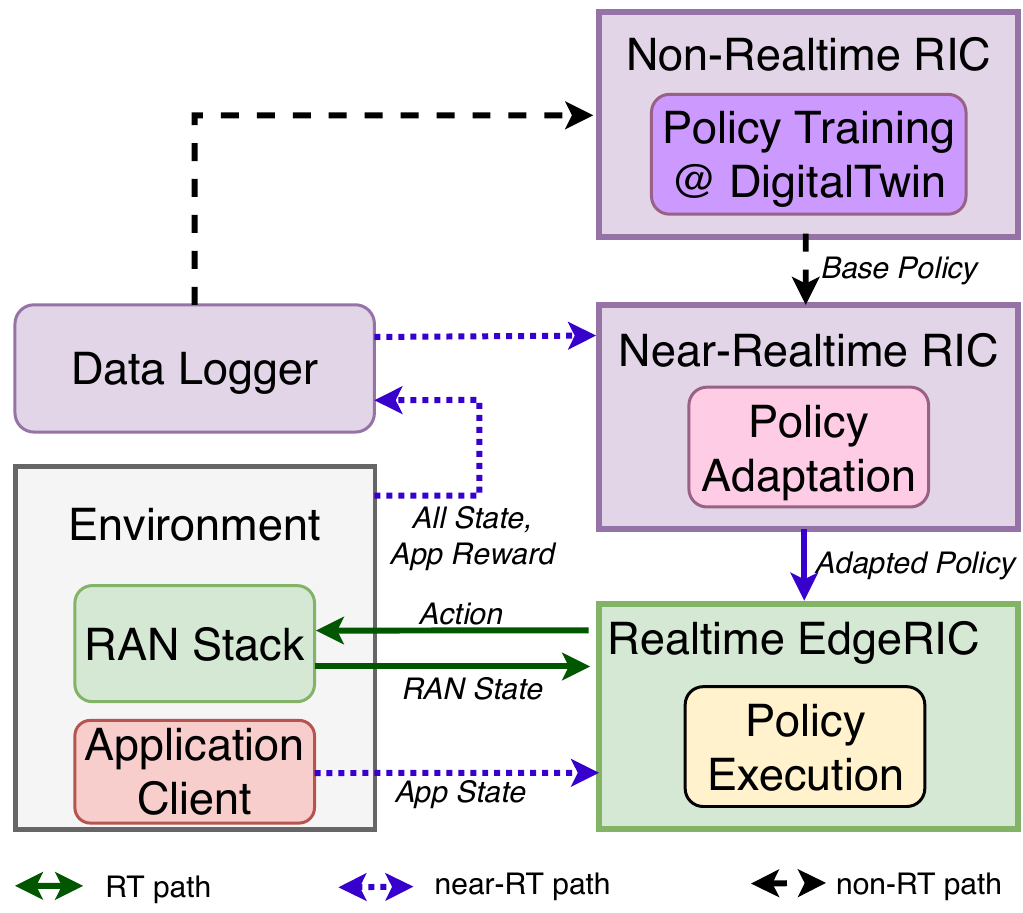}
\caption{\twin\ based Non-RT policy training, Near-RT policy adaptation and RT policy execution. }
\label{fig:workflow}
\end{center}
\vspace{-0.1in}
\end{figure}

The architecture also supports approaches such as Reinforcement Learning (RL), a branch of ML that is explicitly tailored towards learning feedback-control policies.  Here, the core idea is to ``learn by doing'' and to tailor the control policy in an online manner based on the information obtained thus far.  The RL workflow is well aligned with the modality of a \twin\ based non-RT training of a base policy using data collected offline or via an emulator.  Such polices can then undergo near-RT adaptation to the current environment, along with RT policy execution.  Simultaneously, data is gathered at the edge, which is shared with the non and near-RT RIC for accurate training and adaptation.  This three-timescale RL workflow is illustrated in Figure~\ref{fig:workflow}.\looseness-1

\section{EdgeRIC Implementation}
\label{sec:implementation}


We now describe the implementation of EdgeRIC to satisfy the goals of our concept architecture.  The experimental results that we present in this section were collected on two servers: Intel Xeon Gold 5218R CPU @ 2.10GHz, 20 cores and  Intel i9 CPU @ 2.4GHz, 12 cores, without using GPUs.



\subsection{O-RAN Compliant Messaging}\label{sec:messages}

O-RAN specifications provide the E2 interface between the near-RT RIC and the O-DU or O-CU.  Specifically, the E2 application protocol (E2AP) operates over SCTP and provides near-RT services for RAN monitoring and control.  E2 does not support a realtime connectivity service, and we extend the specification to create a realtime varient that we call RT-E2.   RT-E2 supports two basic procedures to connect EdgeRIC with the PHY-MAC stack at the O-DU.

\vspace{0.05in}\noindent \textbf{RT-E2 Report:}  This is a periodic pub-sub procedure under which a module at the O-DU, such as radio link control may publish information at a given rate.  Our default periodicity is one TTI, i.e., information may be generated in realtime.  $\mu$Apps at EdgeRIC may subscribe to the RT-E2 Report service and utilize it for inference and control.   Subscription may be blocking in that the  $\mu$App will proceed only when new information is available from the RAN.

\vspace{0.05in}\noindent \textbf{RT-E2 Policy:}    This is an event-driven pub-sub procedure under which a $\mu$App at EdgeRIC may publish information to one of the O-DU modules  such as UE priorities for resource allocation at the MAC layer.  This information is used directly for realtime control at the O-DU.  Subscription is  non-blocking in that the O-DU subscriber will move on if no new information is available on this procedure, without breaking the tight TTI deadlines required by PHY-MAC.

\vspace{0.05in}\noindent \textbf{TTI-Level Sync Between RAN and EdgeRIC:} It is critical for EdgeRIC to maintain TTI-by-TTI sync between RAN and RT-RIC in order that the control actions sent in realtime from the RT-RIC and the reward information (during training) from RAN accurately correspond to the current state at the RAN.   Hence, we design sync procedures to ensure that RAN and RT-RIC maintain coherence. We maintain TTI counters at the RAN and RT-RIC, called $RANtime$ and $RICtime,$ which increase their counts at the completion of the tasks in that TTI at RAN or RT-RIC, respectively.  

The two possibilities for asynchrony between RAN and RT-RIC and their respective solutions are: (i) \textbf{Lazy RAN:} Here, $RANtime < RICtime,$ which can happen during system initialization.  RT-RIC pauses execution until $RANtime = RICtime$ when a Lazy RAN event happens, (ii) \textbf{Lazy RIC:} Here, $RANtime > RICtime,$ which can happen due to delayed computation of actions at the RT-RIC.  The RT-RIC subscriber always keeps only the latest RAN message in its subscription queue, and sets $RICtime = RANtime$ when a Lazy RIC event happens.  Note that the RAN must have a default option to deal with Lazy RIC events, since it might not receive any control input from the RT-RIC during these times.  
We will demonstrate that with the employment of such sync, the RAN and RT-RIC operate correctly in tandem, and feedback control and RL training proceed successfully.



\subsection{Realtime Connectivity to RAN}

We need a means of decoupling the RAN stack from EdgeRIC, and yet allow realtime information exchange between them.  A message-passing layer can provide such decoupling to ensure that the strictly RT RAN stack is not negatively impacted by the introduction of additional RT components.   We choose ZeroMQ~\cite{zmq} as the inter-process message passing framework over which we implement RT-E2 APIs  for decoupled operation of between RAN and \edgeric.  The low-latency and low-overhead nature of ZeroMQ is well suited for our architecture, which comprises of processes, both with strict RT requirements and looser near-RT requirements.

\begin{figure}[!t]
\begin{center}
\includegraphics[width=\columnwidth]{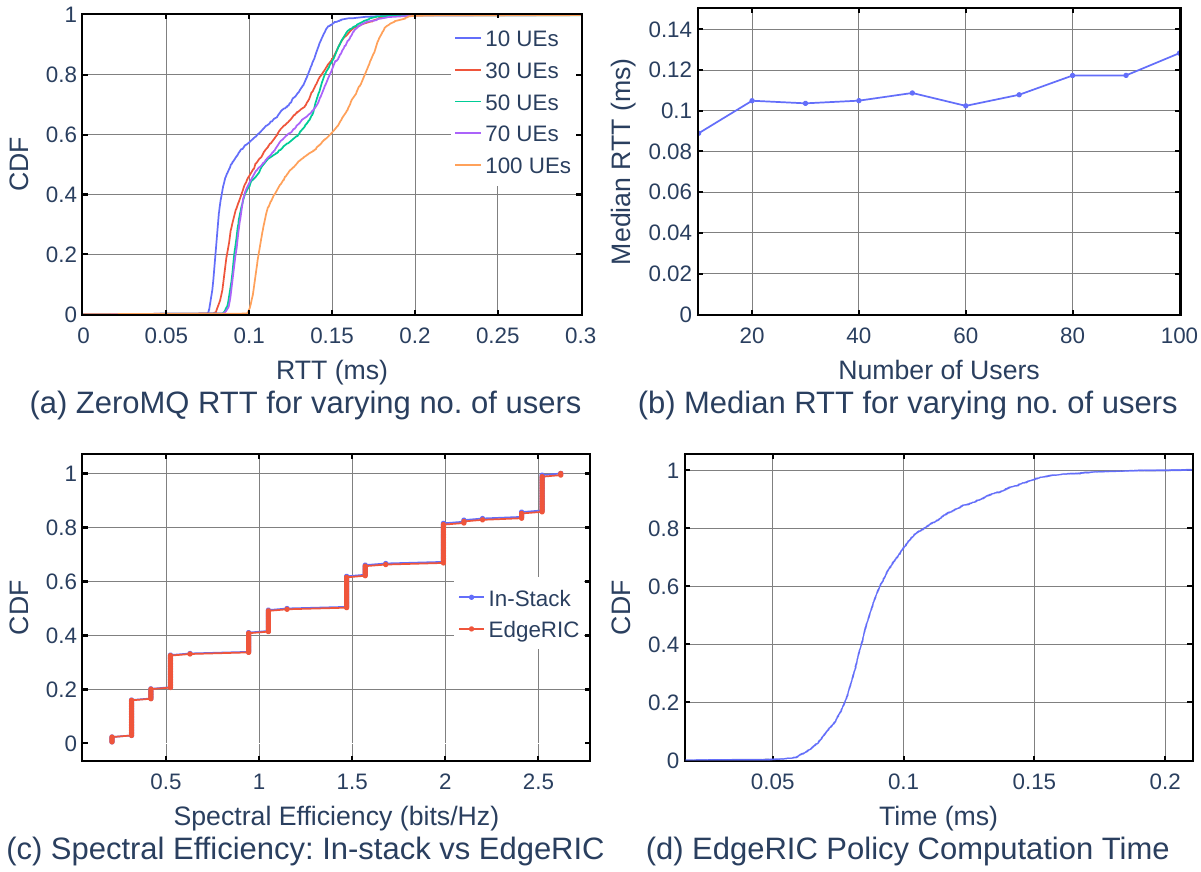}
\caption{\edgeric\ Feedback latency, spectral efficiency and AI-policy execution times. }
\label{fig:cdf_combined} 
\end{center}
\end{figure}

The fundamental timer in our system is the TTI-level clock tick from the RAN stack to which all processes must be synchronized.  We therefore publish RAN-level information once every TTI, and use a RT-E2 Report type blocking subscription with the ZeroMQ conflate option set to `true' (to keep only the latest RAN message in the queue) at RT-RIC  to ensure synchronization of its actions at each TTI.  All other subscriptions are set as non-blocking so that information is obtained whenever it is published, but no action is taken at RT-RIC until the clock tick at the RAN.   Other modules, such as near-RT AI/ML training are self-clocking based on data digests received, with policy updates being published to the RT-RIC. Thus, we permit the coexistence of delay-constrained RT control loops and elastic data streams and policy updates in our architecture.

We conducted an evaluation of the suitability of ZeroMQ for RT operations by running srsRAN stack and the RT-RIC  on a server. The downlink channel operating at 10 MHz was fully loaded, with an over-the-air throughput between physical radios of approximately 37.5 Mbps during these experiments.  The results are shown in Figure~\ref{fig:cdf_combined}(a) and \ref{fig:cdf_combined}(b), which indicate that even while publishing information and receiving control actions for about a hundred UEs (physical and virtual), the median round trip latency of our implementation is an acceptable 100~$\mu$s.  
 In the context of  resource block (RB) allocation at the MAC layer that we focus on as our sample application, this  data acquisition path consists of the RAN stack publishing its state information on the active UEs, including their Radio Network Temporary Identifiers (RNTIs), Channel Quality Indicators (CQIs), backlog buffer states, their past downlink transmit bitrates, and RT-RIC responding with downlink RB allocation decisions for each UE.   We see in Figure~\ref{fig:cdf_combined}(c) that the spectral efficiency of running a policy in EdgeRIC vs. doing so with the same policy embedded within the RAN stack is essentially identical, i.e., the decoupled architecture is successful at maintaining efficiency without compromising on RAN stack robustness.  \looseness-1

\subsection{Cross-Layer Connectivity and Logging}\looseness-1
Our choice of ZMQ for inter-process communication between RAN and EdgeRIC is also extendable to cross-layer application awareness.  Since ZMQ is capable of also operating over TCP or UDP on an IP network, applications can simply use ZMQ to publish their state information to EdgeRIC.  Apart from being lightweight and having APIs in most programming languages, ZMQ also permits client authentication and encryption via CurveZMQ~\cite{czmq}.  This allows secure state-sharing with EdgeRIC.

We also enable \ricsys\  with an in-memory Redis database for data logging and sharing.  Redis is a fast, lightweight, key-value store, in which we log data digests, as well as trained models for sharing across the elements of \ricsys.  An added advantage of Redis is that we can save all traces to drive at experiment conclusion, which allows for post processing and performance analysis.


\subsection{Weight-based Abstraction of Control}\looseness-1
In resource constrained systems, such as wireless radio resource management, optimal policies often have simple and easy to learn structures, such as threshold policies \cite{hsu2014opportunities, mohapatra2014network}, index polices \cite{raghunathan2008index, razavilar2002jointly, parandehgheibi2010access} and linear policies \cite{kumar2015stochastic, bertsekas2017}. In many other cases, the optimal value function may have similar simple and easy to learn structures, such as monotonicity  or concavity \cite{bertsekas2017}. \cite{numfabric} talks about assigning certain weights to prioritize flows, ensuring fairness among flows. All the above structured policies can effectively be represented by assigning relative priorities to the different connected UEs.  For example, the so-called Whittle index is a scalar parameter corresponding to the value of resources allocated to a given UE, which can quickly be learned independently of other UEs~\cite{nakhleh2021neurwin}.  The Whittle indices of the UEs can then be used to prioritize resource block (RB) allocation to those UEs that have high indices.  Resource allocation may also be done with a fairness metric in mind, such as proportional fairness, where RBs are assigned to a UE based on the ratio of its current channel quality as compared to its average channel quality, or max-min fairness~\cite{ShaSri07-Mono}.  Here too, the relative priority of a UE can be represented by a weight assigned to it.\looseness-1

Motivated by the above ideas, our general approach for RB allocation is for EdgeRIC to provide values $w_i$ for each connected UE $i,$ for both uplink and down link over realtime information exchange at each TTI. The 5G MAC will then allocate an number of RBs in a manner proportional to $w_i$ over the next TTI.   Such an abstraction provides simplicity of actions for the resource allocation policy, while maintaining its ability to attain near-optimal allocations in realtime.

\subsection{Integration with OpenAIGym }

 OpenAIGym is an open source python library that provides a framework for developing an interface to interact with and query the environment by any given algorithm.  While it is typically used for developing and comparing reinforcement algorithms, it can be used as a standard approach for realtime policy execution, regardless of whether the policy in question is based on reinforcement learning.  This openness to the nature of the control algorithm motivates us to develop an OpenAIGym interface connecting the RAN stack, EdgeRIC and the control algorithms in the form of $\mu$Apps that it hosts.  This interface involves creating methods that communicate the current state of all the UEs, assign weights to UEs as suggested by the given policy, provide reward feedback for the current allocation, and reset the system when needed. 
 
 Developing the OpenAIGym interface allows for EdgeRIC to be seamlessly integrated with any optimization-based algorithmic approach, such as max-weight scheduling, or using a RL-optimized policy. Thus, the choice of OpenAIGym as our interface allows for swift policy development and freedom of execution of algorithms as desired.  {Figure~\ref{fig:cdf_combined}(d)} shows the time taken by EdgeRIC to execute a forward pass of a trained policy network using only CPU, while running a fully loaded RAN.  The mean value is less than 100 $\mu$s, which implies that TTI-scale execution is straightforward  on general purpose compute.




\section{Deploying AI on srsRAN} 

We chose the open source Software Radio Systems srsRAN stack~\cite{srs} as the experimental RAN system for EdgeRIC integration due to its simple, modular codebase, its stability and compatibility with various core networks, 4G and 5G versions, and the availability of the srsUE codebase.

\subsection{\edgeric\ and srsRAN}\label{sec:edgeric_and_srsRAN}

srsRAN runs on the general-purpose Ubuntu OS, which does not provide realtime guarantees.  Its codebase is optimized to ensure that all fine-grain tasks are completed within the hard-constrained TTI boundaries.  
Thus, integrating of \edgeric\ with srsRAN requires caution to prevent disrupting time-sensitive event chains.   Hence, we adopt a lightweight coupling approach prioritizing two main requirements;

\vspace{0.05in}\textbf{RT-E2 API Support:}  We enable srsRAN codebase modules for radio link control and medium access control to utilize our custom RT-E2 APIs.  These APIs run over ZeroMQ, which is already supported internally in the srsRAN codebase.  Specifically, we use the RT-E2 Report API to collect Radio Network Temporary Identifiers
(RNTIs), Channel Quality Indicators (CQIs), backlog buffer
states, and their past downlink transmit bitrates for each UE, and publish these values each TTI for reception by EdgeRIC.

\vspace{0.05in}\textbf{Weight-based RB Allocation:}  We create a custom scheduler for srsRAN that assigns RBs in proportion to weights provided for each UE.  This scheduler is integrated with the RT-E2 Policy API, and subscribes to weights published by EdgeRIC every TTI in a non-blocking manner.  It verifies correctness of the map between weights and RNTIs  allocates RBs as prescribed.  We maintain reliability by enabling it to use either a default scheduler or fixed weights, in case updates are not received from EdegRIC in a particular TTI.\vspace{0.05in}

\begin{figure}[!t]
\begin{center}
\includegraphics[width=\linewidth]{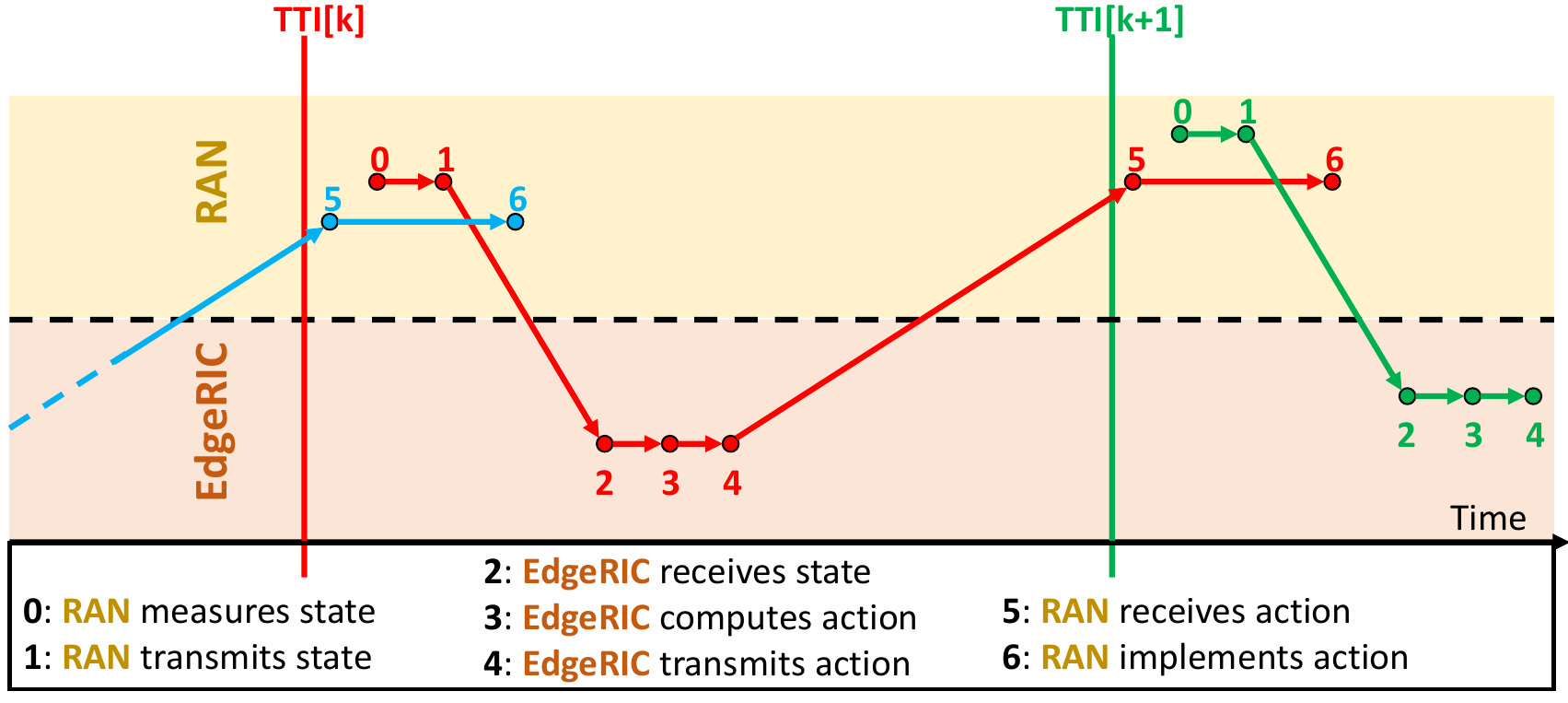}
\caption{TTI-level events for EdgeRIC to RAN loop.  }
\label{fig:tti-level-events}
\end{center}
\end{figure}

\begin{figure}[!t]
\begin{center}
\includegraphics[width=\columnwidth]{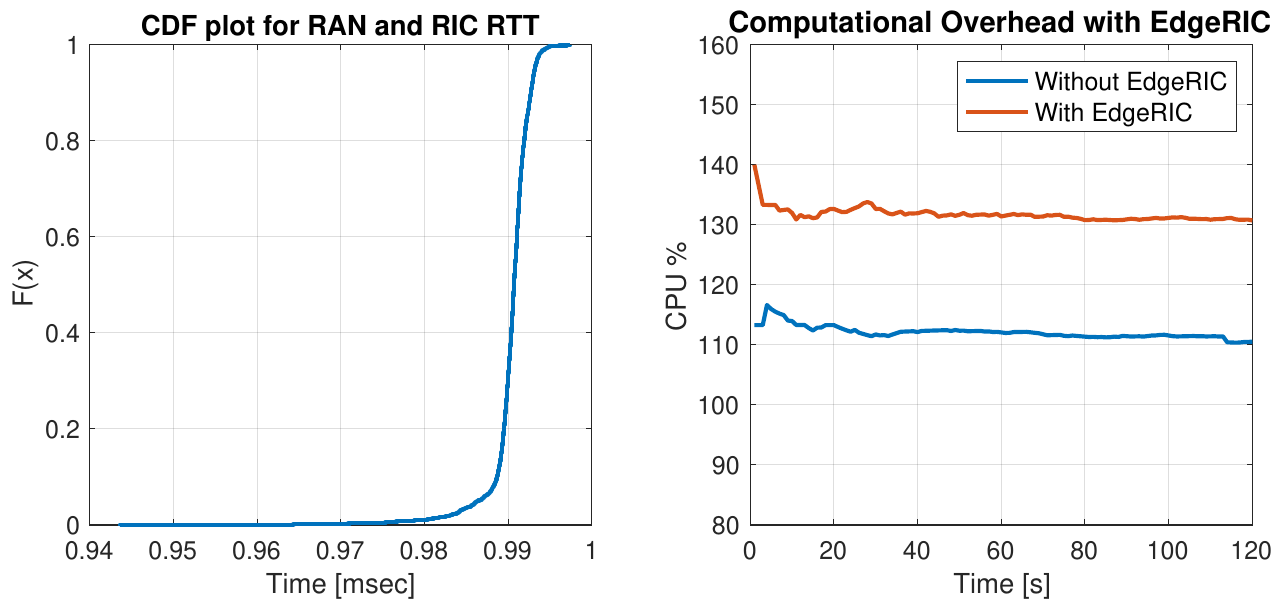}
\caption{CDF of End-to-end RTT between RAN and EdgeRIC, showing that TTI timings are always met.}
\label{fig:cdf_RTT} 
\end{center}
\end{figure}

The full sequence of events is shown in Figure~\ref{fig:tti-level-events}, where we see (i) state measurement and transmission from RAN, (ii) reception, processing and response at EdgeRIC, and (iii) final resource allocation at RAN.  Note that srsRAN receives each EdgeRIC message well before the TTI boundary, but only reads it in a non-blocking manner at the beginning of each TTI.  We show the CDF of end-to-end latency of the entire event chain from RAN to EdgeRIC and back (including policy execution via a forward pass on the policy network), culminating in resource allocation at each TTI in Figure~\ref{fig:cdf_RTT}.  We observe that the end-to-end latency is always less than 1 ms, i.e., the RT-E2 sync procedure 
 (Section~\ref{sec:messages}) successfully meets the target of event completion within each TTI.

Finally, we measure the computational overhead of running \edgeric, as compared to running the vanilla srsRAN stack under a full traffic load.  We see that the difference in CPU utilization is only about 20\%, which means that EdgeRIC is fairly lightweight and does not need execessive additional compute resources.  Hence, co-locating EdgeRIC at the O-DU level seems quite feasible.

\subsection{Implementing \twin}

A major challenge in RL-optimized control is the potential mismatch between the training environment  and the real-world scenario, leading to a gap between the agent's training and real-world performance, known as the "sim-to-real" gap. Constructing a pure Python simulator to model the intricate relationships among the RAN, the RIC, and the end applications is a complex undertaking.
Rather than building a simulator that models each of these components, we leverage virtual radios and channel simulation techniques to train RL policies directly on \twin\ which employs the real RAN and RIC codebases to replicate real-world relationships and performance accurately.  

\begin{figure}[!t]
\vspace{-0.1in}
\begin{center}
\includegraphics[width=\columnwidth]{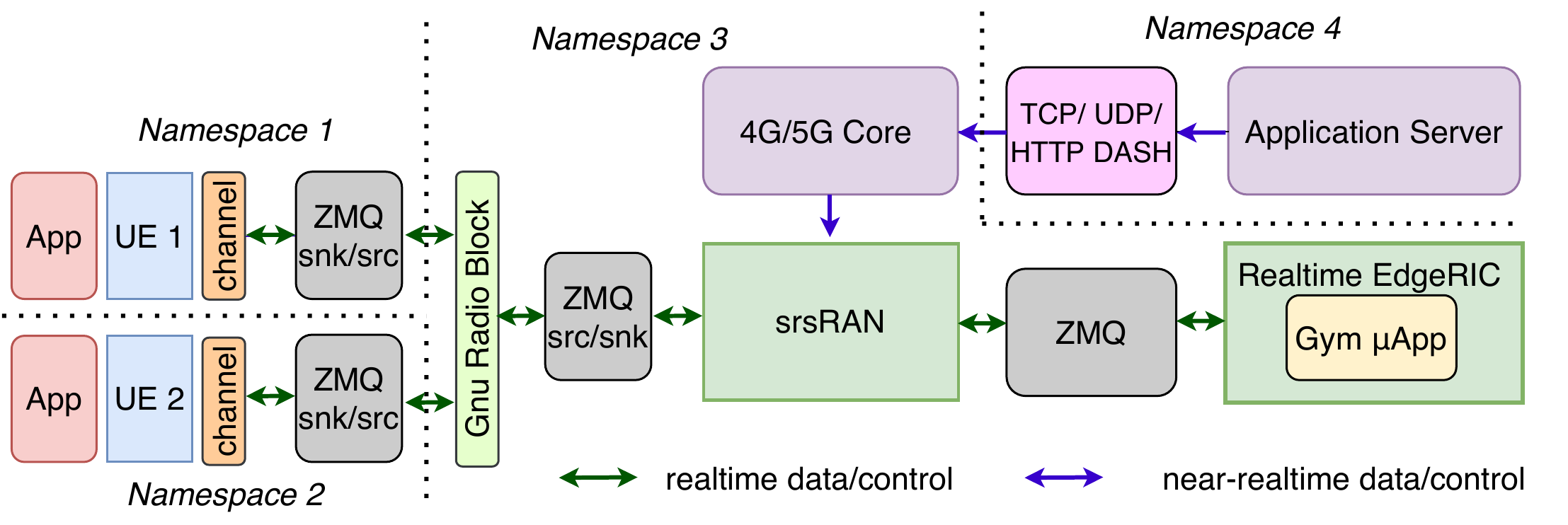}
\caption{\twin\ emulation environment}
\label{fig:digital_twin} 
\end{center}
\end{figure}

The construction of \twin\ is facilitated by existing support for ZeroMQ-based sinks and sources as virtualized radios in srsRAN.
Complex-valued samples, typically sent to a software-defined radio for transmission, are instead sent to a specified ZeroMQ socket.
Additionally, these samples can also be manipulated to introduce simulated channel effects to further bridge the sim-to-real gap of \twin.
We define a GNU Radio flowgraph utilizing the ZMQ Source and ZMQ Sink blocks to multiplex the complex-valued samples to many srsUE inputs, routing them to the appropriate users.
Finally, UEs and application servers can each have their own private IP namespaces, implying that real-world applications using TCP or UDP sockets can be run end-to-end on \twin, as outlined in Figure~\ref{fig:digital_twin}.

Aside from the utilization of virtual radios and a simulated or trace-generated channel, all other components in \twin\ are  clones of the ones used in a real deployment.
Hence, \twin\ has very little sim-to-real gap.

\subsection{Training RL Policies on \twin}
We train an agent to perform optimal resource allocation using the model free RL algorithm, Proximal Policy Optimization (PPO) \cite{ppo}. We chose PPO due to it's ease of implementation and high sample efficiency.
An iteration of training consists of collecting 5000 samples from the environment, adjustment of the agent's policy neural network weights through backpropagation, and the utilization of the updated agent to generate an additional 5000 samples to assess its performance and chart the training curve. A sample represents one transmission time interval (TTI) in the real system and consists of - the current state of the environment, action taken by the agent in this state, and the reward and next state observed as a result of this action.  We may train RL policies on \twin\ for a variety of use cases, which can utilize either RAN information alone, or be augmented with application-level information for cross-layer optimization.  Specially, we will focus on two use-cases on (i) downlink throughput maximization, discussed below and in Section~\ref{sec:microbenchmark}, and (ii) video streaming stall minimization, discussed in Section~\ref{sec:videostreaming}.

With the a throughout-maximization objective, we choose the RAN-level state information as CQI and backlog buffer lengths for each UE, while the action is the allocation weights mentioned in Section~\ref{sec:edgeric_and_srsRAN}, and the reward is the total throughput achieved.  Reward saturation for this use-case was typically observed under 100 iterations of training, which equates to 500,000 training samples (TTIs). The RL framework specifications for each this setup is presented in Table~\ref{tab:RLspec_MicroBenchmark}.  Here,  $CQI_i[t]$ denotes the channel quality index of UE $i$ at time $t,$ $B_i[t]$ denotes the number of backlogged bytes in the downlink queue for UE $i$ at time $t,$ and the action returned is the weight $w_i[t]$ accorded to UE $i$ at time $t$.    

Including data collection/transfer overheads to the RL agent, and the time taken to perform actor-critic policy update at the end of each iteration, the total time taken for training completion in \twin\ is within approximately ten minutes.  We present training results for this use-case on \twin\ below, while experiments on execution over \edgeric\ are shown in Section~\ref{sec:microbenchmark}.

\begin{table}
\caption{RL Specifications: Throughput Maximization}
\centering
\begin{tblr}{
  hlines,
  vlines,
}
State $(s[t])$ & $B_i[t] , CQI_i[t] \; \forall i$ \\
Action $(a[t])$ & $w_i[t]     \;  \forall i$                  \\
Reward $(r[t])$ & \textit{total throughput}              
\end{tblr}
\label{tab:RLspec_MicroBenchmark}
\end{table}

\subsection{Case Studies on \twin }
\label{sec:base-algos}
 We conducted emulations using synthetic channel traces in \twin to assess the potential gains achievable by a real-time agent for policy computation and control. Two metrics we use throughout our study is the downlink system throughput and the downlink backlog buffer lengths at the RAN. We desire to evaluate two basic questions, \emph{(i)~How much does performance improve with realtime control as opposed to near real time control} and \emph{(ii)~How does RL-based control policy perform compared to basic algorithms?}

 We consider three basic algorithms for weight-based resource allocation.  In all the below approaches, each UE is assigned a weight $w_i[t]$ at TTI $t.$  The weights are then normalized over all UEs as $\tilde{w}_i[t] = w_i[t]/\sum_j w_j[t].$   The RAN receives the normalized weights $\tilde{w}_i[t]$ from the RIC, and  performs an allocation of resource block groups (RBGs) in proportion to the weights, i.e., $R_i[t] = \tilde{w}_i[t] R_{total}[t],$ where $R_i[t]$ is the assignment to UE $i,$ and $R_{total}[t]$ is the number of RBGs available in TTI $t.$  While some approaches call for an absolute prioritization of UEs that have a maximum weight~\cite{tassiulas1990stability}, we find in practice that a proportional division based on weight leads to better overall performances.

\noindent\textbf{CQI-Fair Allocation:}  Here, the weight of UE is equal to its realized CQI.  Hence, $w_i[t] = CQI_i [t],$ where $CQI_i[t]$ is the realized CQI of UE $i$ at time $t.$   This approach effectively tries to obtain a large total throughput by prioritizing these UEs that have a large CQI in the current timeslot.

\noindent\textbf{Proportionally-Fair Allocation:} Here, the weight  of UE is the ratio between its current CQI and its average CQI, with the idea of prioritizing those UEs that have a good channel realization compared to their average value.  The average CQI, denoted $AvgCQI_i[t]$ for UE $i$ is calculated using an exponentially weighted moving average for each UE up to the current time, $t$.  Thus, we have, $w_i[t]=CQI_i[t]/ AvgCQI_i[t].$ 

\noindent\textbf{Max-weight Allocation:}  Here, the weight of a UE is the product of its current CQI and the backlogged bytes in the downlink queue corresponding to that UE.   The max-weight policy is known to be throughput optimal~\cite{tassiulas1990stability}, in that it can achieve the capacity region of the system.  Thus, we have, $w_i[t] = CQI_i[t] B_i[t],$ where $B_i[t]$ is the number of backlogged bytes in the downlink queue of UE $i.$

To answer the first question, the performance of downlink resource block (RB) allocation algorithms operating as $\mu$Apps on EdgeRIC is compared to their operation as xApps on a cloud-based RIC, the distinction being the latency in state capture and control. A $\mu$App at the edge experiences a low round-trip latency of a few tens of microseconds between receiving state information from the RAN and action generation, whereas an xApp in the cloud experiences both forward and reverse network latency between the RAN and the cloud, resulting in a round-trip time of tens of milliseconds. To demonstrate the effect of cloud latency, appropriate delays are artificially induced in \twin.

\begin{figure}[!t]
\vspace{-0.1in}
\begin{center}
\includegraphics[width=\columnwidth]{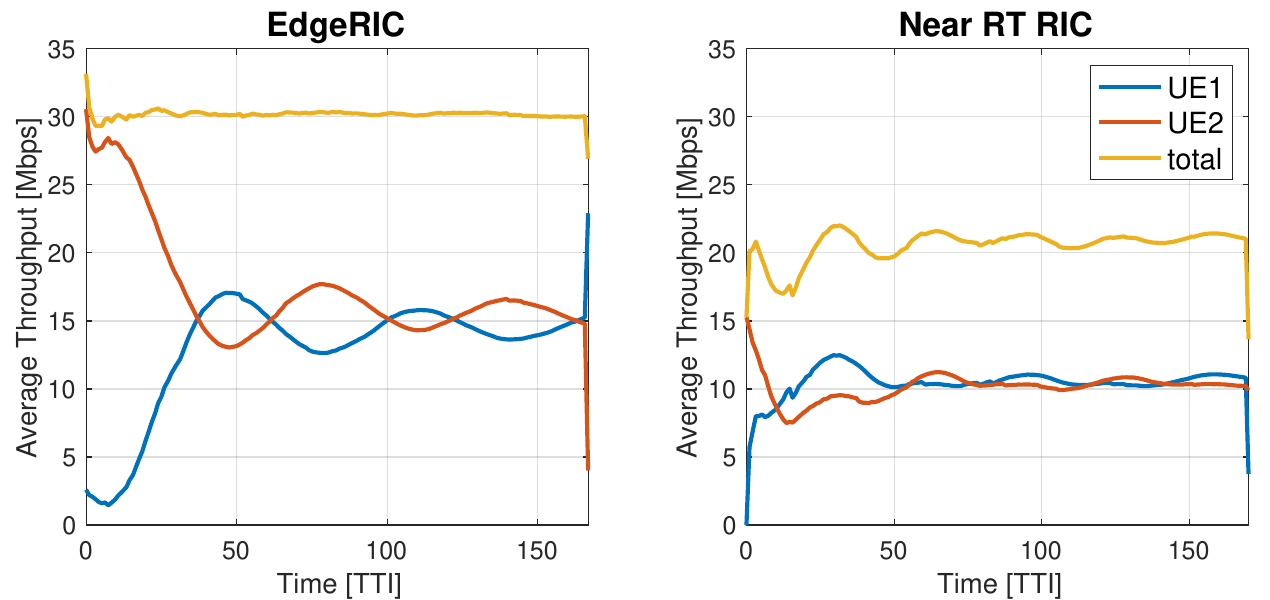}
\caption{CQI-Fair Allocation with 2UE synthetic channels - one can support higher throughput while maintaining the stability of backlog buffers}
\label{fig:cqi_fair} 
\end{center}
\end{figure}

We begin our experiment with synthetic channel traces.  The results of the CQI-Fair allocation algorithm, when executed as a $\mu$App over EdgeRIC or as a Cloud-based RIC with a total monitoring and control latency of 30ms, are depicted in Figure \ref{fig:cqi_fair}. The findings suggest that the utilization of EdgeRIC leads to a 50\% increase in throughput while preserving the stability of backlog buffers. This superiority in throughput performance is further illustrated in Figure \ref{fig:4ue_standard} in the appendix, which portrays the results of a 4-user scenario. The results are summarized in Tables \ref{2UE synthetic}, \ref{4UE synthetic}. We see that real-time control significantly improves performance metrics in most cases.\looseness-1

\looseness-1

\begin{figure}[!t]
\begin{center}
\includegraphics[width=\columnwidth]{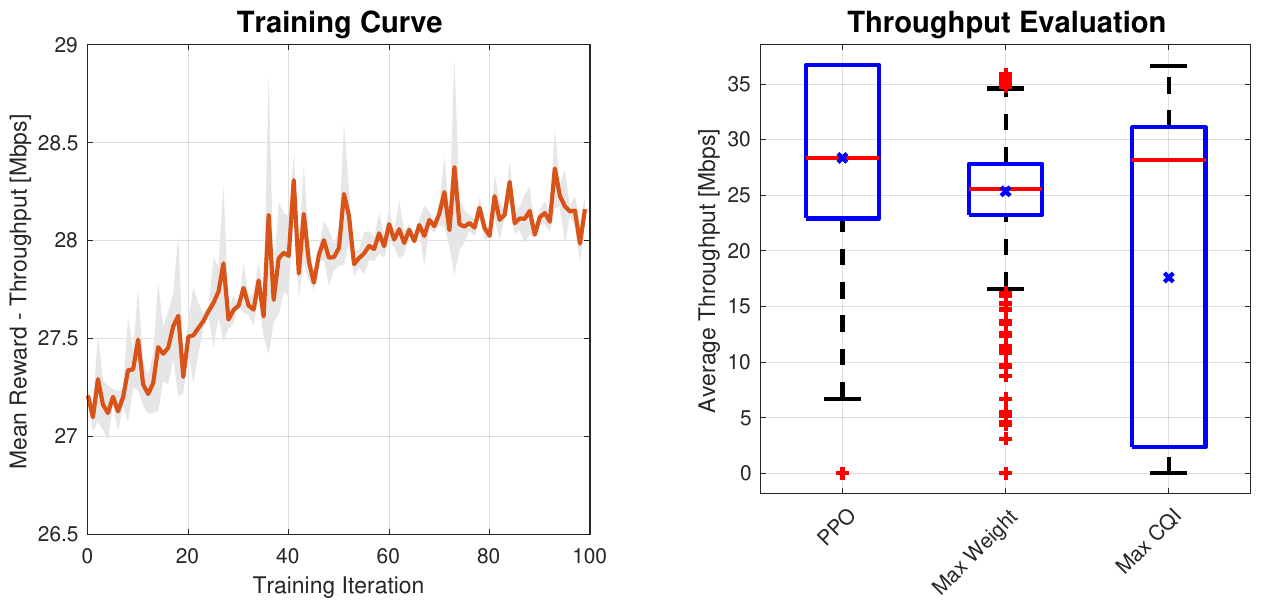}
\caption{RL training and evaluation on a synthetic channel trace}
\label{fig:digitaltwin_rl} 
\end{center}
\vspace{-0.1in}
\end{figure}

To answer second question, we demonstrate the training and evaluation of an RL algorithm on the \twin. Scenario 1 is based on a 2UE environment.
both connected users have uniform variation in CQI values over time, ranging from 1 to 15. 
In scenario 2, one user experiences good channel conditions (CQI values between 8 and 15), while the other user experiences poor channel conditions (CQI values between 1 and 7). In Scenario 3, the CQI values are randomly generated. Figure \ref{fig:digitaltwin_rl} shows training and evaluation on Scenario 2 and Table \ref{tab:RLevaluation_synth} summarizes the performance of RL algorithms on synthetic channel traces. \looseness-1

\begin{table}[!t]
\caption{Load: 35Mbps, Channel: 2 UE synthetic channel}
    \centering
    \begin{tabular}{|l l c c c|} 
 \hline
  &  & EdgeRIC & 15ms & 30ms \\ 
 \hline
 Max CQI & Avg. Thrpt. & \textbf{32.6} & 24.2&18.0 \\ 
 \hline
  & BL[MB] & 0.61 & 0.64&\textbf{0.57} \\
 \hline
 
 Prop. Fair. & Avg. Thrpt. & \textbf{30.7} & 25.7& 21.9 \\ 
 \hline
  & BL[MB] & \textbf{0.65} & 0.67&0.68 \\
 \hline
 
 Max Weight & Avg. Thrpt. & \textbf{30.0} & 23.3& 20.9 \\ 
 \hline
  & BL[MB] & \textbf{0.60} & 0.62&0.65 \\
 \hline

\end{tabular}
\vspace{0.1in}
\label{2UE synthetic}
\end{table}  

\section{\ricsys\ Evaluation}

In this section, we evaluate \edgeric\ by addressing three fundamental questions. These questions are: \emph{(i) Does real-time RAN control through $\mu$Apps on \edgeric\ outperform near-real-time control through a cloud-based RIC, (ii) Is it feasible to implement real-world AI optimization (RL training and feedback) for resource allocation over \edgeric\ }, and \emph{(iii) Does application state feedback to \edgeric\ provide significant improvement to the end-user Quality of Experience (QoE)}?


To evaluate the above questions, we first present our base-station setup and four types of UEs, from fixed to highly mobile users. Next, we discuss a variety of experiments conducted to result in dynamic wireless channels. We used these channel traces to evaluate the performance of traditional schedulers and an RL-based scheduler run as $\mu$Apps on \edgeric\ . We also present results from real-world over the air experiments to support our claims. Finally, we provide a case study of a streaming application that  benefits from this platform. Our RL-based scheduling algorithm outperformed traditional methods, confirming EdgeRIC's capabilities.




\subsection{Experimental Setup}
The EdgeRIC implementation is a modular extension to the srsRAN codebase, which can accommodate both software-defined radios and off-the-shelf UEs. Our base-station stack was built using srsRAN version 21.10 to support EdgeRIC, and we utilized USRP B210 SDRs, an edge DU, and an embedded delay of 20ms in the CU stack. Our srsRAN base-station includes a logging mode for channel trace logging. To avoid interference and to move freely without disrupting building WiFi or commercial cellular transmissions, our experiments were conducted within the 2.5 GHz EBS band under experimental FCC license.  Our typical experiments were with 10 MHz bandwidth, with a fixed base station and mobile users.

\subsubsection{User Devices and Channel Traces} 
\begin{figure*}[!t]
\begin{center}
\includegraphics[width=\linewidth]{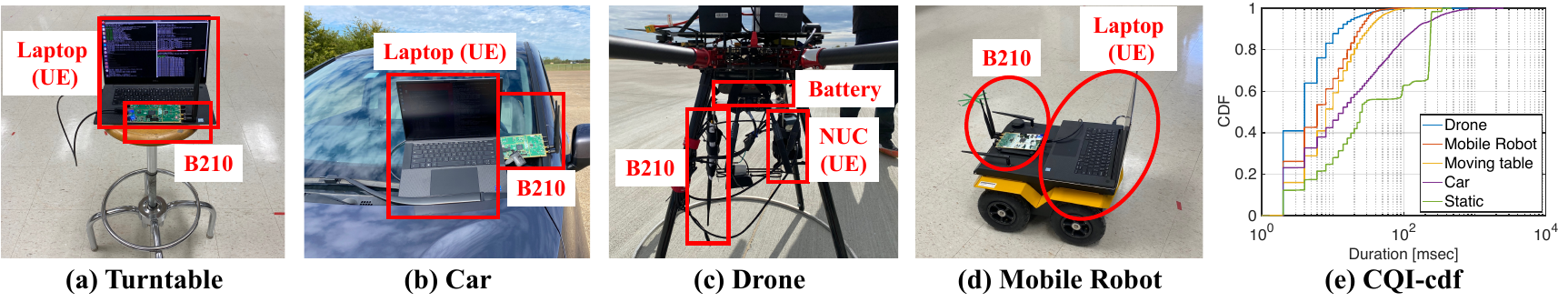}
\vspace{-0.1in}
\caption{Experimental setups for collecting CQI data of various mobility and CDFs of their CQI traces}
\label{fig:exp_setup} 
\end{center}
\vspace{-0.1in}
\end{figure*}

We use a static base-station with different UE types to collect channel traces as shown in Fig.~\ref{fig:exp_setup}. We considered various mobility models, both within the laboratory and in outdoor environments. We extract the channel quality indicator (CQI) values sampled at each transmission time interval (TTI) for each user at the srsRAN for runs of approximately 8 minutes each.  Next, we summarize the UEs considered in this study.


\textbf{TurnTable UE:} Figure \ref{fig:exp_setup}(
a) shows the setup of a UE on a movable turntable, similar to a user sitting on a chair and rotating. The base station equipped with a B210 is on a static table, while a UE node with a B210 is mounted on the turntable. The UE node is moved away from or towards the base station at about 1 m/s and rotated to produce significant variations in its CQI values. The distance between the UE node and the base station ranges from 0.5 meter to 4 meter.

\textbf{Car UE:} Figure \ref{fig:exp_setup}(b) shows a UE setup in a car, using a USRP B210 and laptop powered by the car. CQI data was collected while driving along three paths: 1) a 6-meter-radius circle with a maximum speed of 4.5 m/s, 2) a 30 meter x 3 meter rectangle with a maximum speed of 5.4 m/s, and 3) a 50-meter straight line with a maximum speed of 9 m/s. 

\textbf{Drone UE:} We used a drone experiment to demonstrate performance with faster channel variations in 3D space. Figure \ref{fig:exp_setup}(c) shows a B210 and a NUC (a small and portable computer) mounted on a Big-Hexy hexa-copter drone, while the base station, equipped with another B210, was placed on the ground. The drone was flown over the base station along two paths: 1) a 10-meter straight line at a height of 15 meters and with a speed of 2.2 m/s, and 2) a 15-meter straight line at a height of 20 meters and with a speed of 3 m/s.

\textbf{Indoor robotic UE:} Mobile robots were used to for indoor mobility experiments. In Figure \ref{fig:exp_setup}(d), a B210 and laptop were mounted on a Jackal robot, while the base station with an X310 was placed on a table. The robot moved along a 1.6 m x 1.6 m square path and rotated at each corner, causing CQI values to drop due to signal blockage. The working area was limited to 4 m from the base station, similar to robot control or industrial IoT with Private 5G.  

We summarize the results in Figure \ref{fig:exp_setup}(e), which shows the CDF of the time interval between CQI changes.  We see that several mobility scenarios yield CQI changes in the sub 10 ms range, with the drone trace showing this effect for almost 90\% of samples.
\vspace{-0.3 cm}
\subsubsection{Evaluation Scenarios}

To generate realistic scenarios, we collected channel traces from various environments and utilized them in our DigitalTwin emulation setup. This setup includes the same core, radio access network (RAN), and user equipment (UE) modules as our over-the-air experiments that generated the CQI traces. By replaying the CQI traces with a desired number of UEs, algorithmic methods can be compared while maintaining the same end-to-end applications and channel conditions. 
For end-to-end over-the-air experiments, we used a X310 as a base station and two B210s as UEs.  One of the UEs had lower channel quality than the other by placing them at different distances from the base station. 
Table~\ref{tab:scenarios} summarizes the various scenarios considered in this study.

\begin{table}[]
\caption{Summary of all scenarios}
\centering
\begin{tabular}{|l |l|}
\hline
Scenario & Channel Description \\
\hline
\multicolumn{2}{|l|}{Channel Traces from Experiments} \\
\hline
Scenario 1& 2 Drone UEs\\
\hline
Scenario 2 &  2 Turntable UEs\\
\hline
Scenario 3& 2 Car UEs and 2 Drone UEs \\
\hline
Scenario 4 &  2 Car UEs and 2 Indoor Robotic UEs \\
\hline
Scenario 5 &  2 Random Walk UEs and 2 Turntable UEs\\
\hline
\multicolumn{2}{|l|}{Complete Over-the-Air Experiments} \\
\hline
Scenario 6 & 2 UEs on indoor mobile robots \\
\hline
Scenario 7 & 2 UEs on indoor stationary robots \\
\hline
\end{tabular}
\label{tab:scenarios}
\vspace{-0.1in}
\end{table}

\subsection{Micro-benchmarks: Edge vs. Cloud}

We utilize iPerf for measuring network throughput and generate microbenchmarks. We test our scenarios on different UEs, each with varying traffic loads, and report the throughput and backlog buffers observed with different scheduling algorithms presented in Section~\ref{sec:base-algos}.

\begin{table}[]
\caption{Load: 35Mbps, Channel Trace: 4 Turntable UEs}
    \centering
    
    \begin{tabular}{|l l c c c|} 
 \hline
  &  & EdgeRIC & 50ms & 100ms \\ 
 \hline
 Max CQI & Avg. Thrpt. & \textbf{33.4} & 21.2&29.5 \\ 
 \hline
  & BL[MB] & 1.34 &\textbf{0.84}&1.12 \\
 \hline
 
 Prop. Fair. & Avg. Thrpt. & \textbf{28.6} & 26.6&23.5 \\ 
 \hline
  & BL[MB] & 1.20 & 1.29&\textbf{0.93} \\
 \hline
 
 Max Weight & Avg. Thrpt. & \textbf{33.2} & 28.8&31.0 \\ 
 \hline
  & BL[MB] & 1.14 & 1.30 & \textbf{1.12} \\
 \hline
 
\end{tabular}
\label{4UE_realistic}
\end{table}

\begin{figure*}[]
\begin{center}
\includegraphics[width=\linewidth]{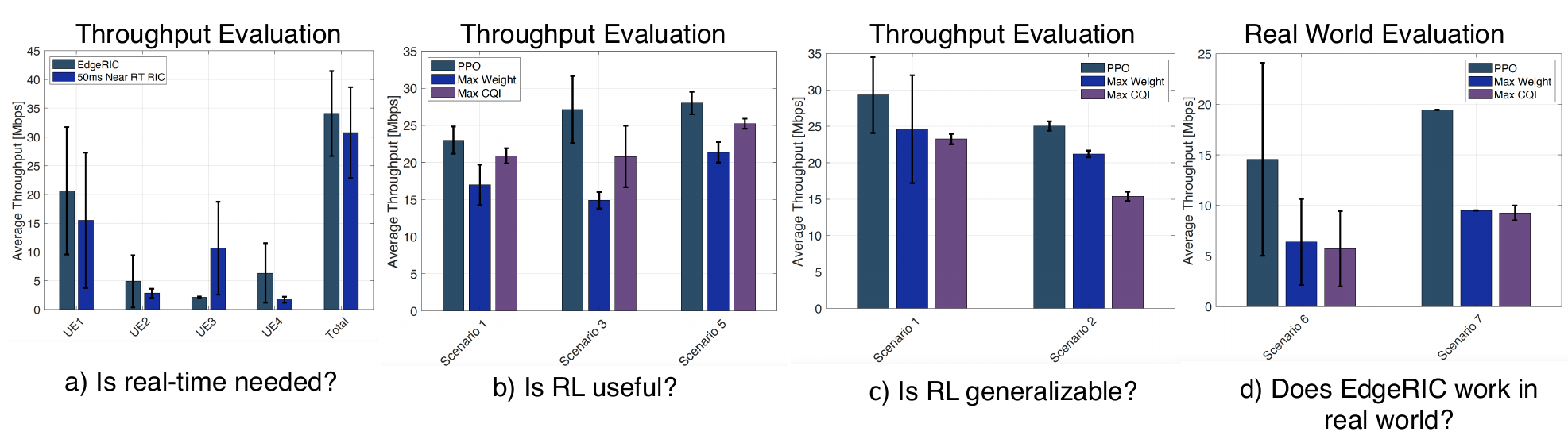}
\vspace{-0.4in}
\caption{Snapshot of EdgeRIC performance}
\label{fig:evaluation} 
\end{center}
\vspace{-0.1in}
\end{figure*}

The validity of our results presented in Section 5.4 is confirmed using realistic channel traces (4 Turntable UEs), as depicted in Table \ref{4UE_realistic} Figure \ref{fig:evaluation}(a). This provides robust evidence in support of the hypothesis that real-time control can significantly improve the system throughput.

\subsection{Impact of RL on Micro-benchmarks}\label{sec:microbenchmark}

We present compelling evidence for the feasibility, impact, and robustness of RL in executing real-time control policies. We begin by addressing the question of RL training in a digital twin environment, which is answered by the set of curves in Figure \ref{fig:RLtraining} of the appendix. These curves demonstrate that RL training with realistic channel traces is highly efficient, with an agent typically converging in just 20 to 40 iterations. The potential of RL is evident from a series of figures that answer important questions about its application.  

Figure \ref{fig:evaluation}(b) shows us that RL can achieve higher throughput than traditional algorithms, bringing us to answer: is RL truly useful? This is further summarized in Table \ref{tab:RLevaluation}, which displays the total system throughput and the mean of the total backlog buffer for various scenarios, offering a snapshot of the RL performance in real-time. The values displayed in the table represent the throughput in Mbps (left) and the backlog buffer in MBytes (right).

Figure \ref{fig:evaluation}(c) offers an answer to the question of model transferability.  It demonstrates that an RL model robustly trained in one scenario can be effectively used in another, revealing its generalizability to other scenarios with similar numbers of connected users. A model trained on the random walk scenario was used to evaluate on a 2 Drone UEs scenario (Scenario 1) and on a 2 Car UEs and 2 Turntable UEs scenario (Scenario 2). Yet, the third question is arguably the most important: can RL really provide performance gains with realtime wireless control in the real-world? Figure \ref{fig:evaluation}(d) offers a clearly affirmative answer. While conducting real-world experiments with EdgeRIC, we observed that excessive channel variations led to unreliable performance of traditional algorithms like max weight and max cqi, despite their theoretical effectiveness. This resulted in limited data transfer. However, the robustness of the RL policy ensured system stability and delivered a throughput that matched the load.



\begin{table}[]
\caption{Throughput and Backlog Buffer Evaluation}
\centering
\begin{tabular}{l c c c }
\hline
 & PPO & Max Weight & Max CQI \\
\hline
\multicolumn{4}{l}{Realistic Channel Traces} \\
\hline
Scenario 1 & \textbf{29.1/0.38} & 26.1/0.53 & 14.9/0.39  \\
Scenario 2& 30.5/\textbf{0.38} & \textbf{31.9}/0.43 & 14.42/0.39  \\
Scenario 3& \textbf{25.3}/1.5 & 22.9/1.3 & 18.67/\textbf{0.97}  \\
Scenario 4& \textbf{25.9}/1.5 & 23.9/1.21 & 20.3/\textbf{1.05} \\
Scenario 5& \textbf{28.5/0.96}  & 26.3/1.46 & 23.3/1.01  \\

\hline
\multicolumn{4}{l}{Over the Air Experiments} \\
\hline

Scenario 6& \textbf{14.6/0.19} & 6.4/0.45 & 5.7/0.44  \\
Scenario 7& \textbf{19.33/0.05} & 10.71/0.34 & 9.06/0.35  \\
\hline
\end{tabular}
\label{tab:RLevaluation}
\end{table}

\subsection{Cross-Layer Optimization: Case Study}\label{sec:videostreaming}
We next consider the potential for enhancing application performance via cross-layer optimizations in a video streaming case study. To this end, we designed a heterogeneous application environment comprising four users. Two are linked to a HTTP video server, utilizing a high-quality video streaming application with a constant bit rate, while the others act as system loaders, simulating a background file downloader via iPerf traffic with 10 Mbps system load. 

Our study uses the open-source GPAC \cite{gpac} library as the video streaming platform, generating 2-second DASH-style segments at the video server, with a media buffer size of 6 seconds at the client end. The video frames display at a rate of 24 frames per second.  When the media buffer falls below 2 seconds, it is considered a stall event.  The stall duration is computed until the buffer refills to over 2 seconds and is ready for playback again. We updated the media buffer size information at the client every time a frame is about to start, every 40 milliseconds in a 24fps video, and published it to EdgeRIC. This allowed the EdgeRIC learning agent to be informed of the application state in the form of the media buffer size every 40 ms, along with network states at each TTI.  To emulate a realistic scenario, we introduced a delay of around 20ms to simulate the uplink latency on srsRAN, as the media buffer size information needs to traverse through the core to the server before it becomes available to EdgeRIC. 

 \begin{figure}[htbp]
\begin{center}
\includegraphics[width=\linewidth]{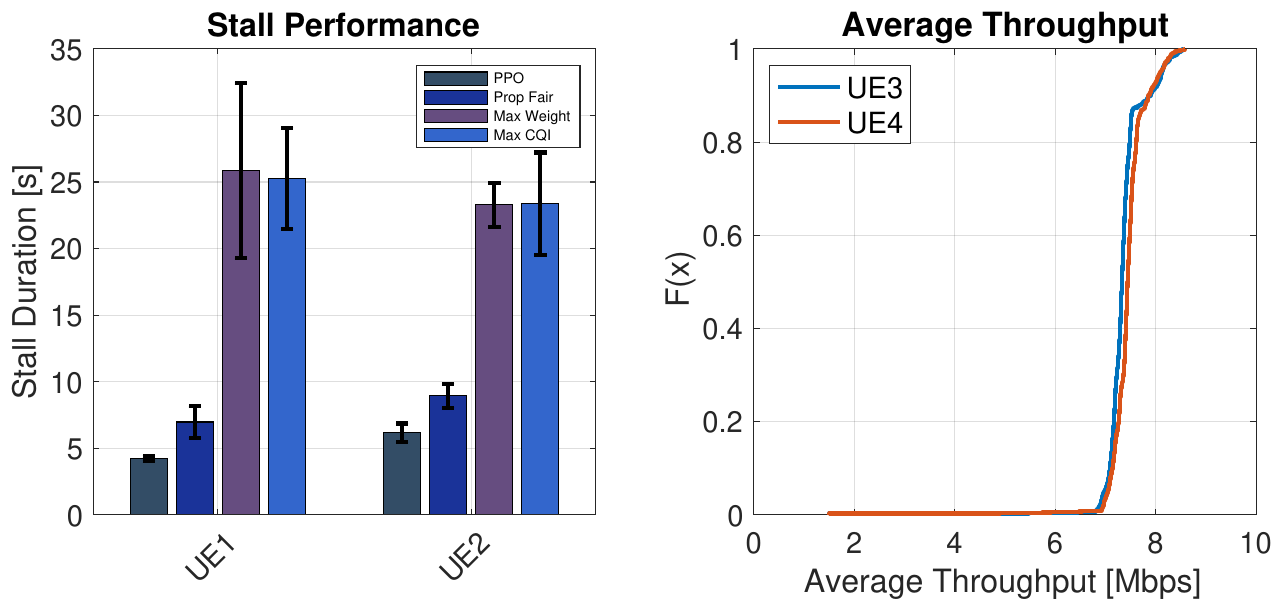}
\vspace{-0.2in}
\caption{Application Performance}
\label{fig:macrobenchmark} 
\end{center}
\vspace{-0.1in}
\end{figure}

\begin{table}
\caption{RL Specifications: Video Streaming}
\centering
\begin{tblr}{
  hlines,
  vlines,
}
State $(s[t])$ & $B_i[t] , CQI_i[t], MB_i[t] \;  \forall i$ \\
Action $(a[t])$ & $w_i[t]     \;  \forall i$                  \\
Reward $(\sum_{i}r_i[t])$ & \(\displaystyle r_i[t]=\begin{cases}
-20,& \text{if } MB_i[t] < 2\ sec\\
+2,              & \text{otherwise}
\end{cases}
\; \forall i
\)           
\end{tblr}
\label{tab:RLspec_VideoStreaming}
\end{table}

An RL agent is trained on \twin\ to execute realtime policies for a video streaming application in a heterogeneous environment with EdgeRIC, guided by a reward function that aims to achieve optimal video playout without stalling and efficient resource allocation to system loaders.  The RL framework specifications for this setup are presented in Table~\ref{tab:RLspec_VideoStreaming}. Here, apart from downlink backlog $B_i[t],$ and channel quality $CQI_i[t]$ of UE $i,$ we augment the state with the length of the media buffer (in seconds) of the video streaming application of UE $i,$ denoted $MB_i[t].$  The reward now depends on the stall performance of the applications, with smooth playout receiving a positive reward, and a stall receiving a large negative reward. We found after the training process that the RL agent rapidly converged to an optimal policy within ten iterations. The RL policy effectively allocated sufficient resources to UE3 and UE4, with an average throughput of about 7.5 Mbps per user, maintaining a stable backlog buffer. 

We next conduct real-world experiments on the video streaming application.  Comparing the video streaming performance of UE1 and UE2 controlled by the RL policy and the standard RAN scheduling algorithms in real-time, the RL policy outperformed the standard algorithms by incorporating "application awareness," demonstrating its potential to provide a quality of experience (QoE)-optimized solution with proper training. Figure \ref{fig:macrobenchmark} summarizes this statement showing the metrics observed by playing a video for 120s. Table \ref{tab:realworld_video} presents the results of our experiment, which evaluated the stall performance of the video streaming application over the air in a static environment with a one-streamer and one-loader scenario for 30 s videos.  We see that the RL trained policy only has about a third of the stalls experienced by the other approaches.

\begin{table}[]
\caption{Total Stall Duration: Video Streaming with EdgeRIC in real world experiments}
\centering
\begin{tabular}{l c c c }
\hline
 & PPO & Prop. Fairness & Max Weight \\
\hline
Run 1 & 2.6s& 4.2s & 5.4s \\
\hline
Run 2 & 2.3s & 3.8s & 4.5s \\
\hline
\end{tabular}
\label{tab:realworld_video}
\end{table}

\section{Limitations and Future Work}
\label{seconclusion}

We presented a platform \ricsys, with a realtime controller microservice, \edgeric and a full stack emulator, \twin, which, when combined together, can train deploy cross-layer AI-optimization algorithms that can provide decision and control at a millisecond scale.


\noindent\textbf{Extension across different ORAN stacks:} Our work has demonstrated that simple modification to srsRAN (open source stack) enables us to support \edgeric. Going forward, will open-source our implementation for the community to re-create the results and scale it to different black box and white box implementations of the RAN stack for deployment. 

\noindent\textbf{Extension to different applications:} We are probably the first to show end-to-end realtime over-the-air experiments that demonstrate the feasibility and value of RL-based policies operating over a 5G stack.  We do acknowledge our system needs testing with multiple applications, and open-sourcing the platform will help community efforts  to do so. 


\noindent\textbf{Scale to 100s of users:} Our demonstrations are with four users, primarily due to the instability with our setup of srsRAN for a larger number of UEs.  However, we have shown that  \edgeric platform can send the messages for 100 UEs, with a RAN to EdgeRIC latency of about 100$\mu$s (Figure \ref{fig:cdf_combined}).   That said, we have not tested over-the-air with support for many more simultaneous UEs, which is key future work. 


\noindent\textbf{Mutliple base-stations:} Our work has demonstrated performance improvements for a single base-station. We anticipate with multiple base-stations, we would have an EdgeRIC attached to each base-station/DU, with Near-RT RIC coordinating all the EdgeRIC instantiations.  This would involve problems in federated learning that we will explore.

\noindent\textbf{Ethical concerns:} This work does not raise any ethical issues.









\bibliographystyle{ACM-Reference-Format}
\bibliography{references}
\appendix
\section{Appendix}

In the appendix, we provide additional results for different scenarios, which are not critical to the performance but provide visibility into the system. 

\begin{figure}[htbp]
\vspace{-0.1in}
\begin{center}
\includegraphics[width=\columnwidth]{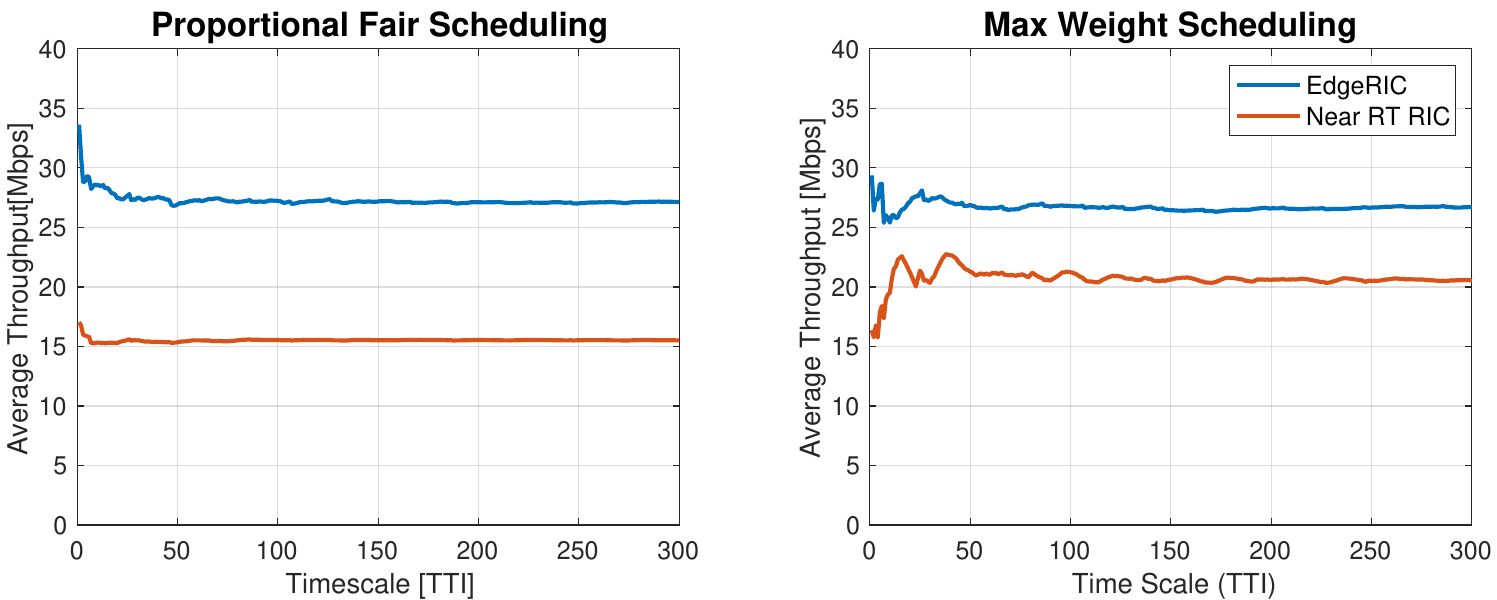}
\caption{EdgeRIC Performance in a 4UE synthetic channel scenario.}
\label{fig:4ue_standard} 
\end{center}
\end{figure}

Figure \ref{fig:4ue_standard} shows that EdgeRIC outperforms Near Real-time RIC for implementation of microbenchmarks. This is the scenario with 4UEs on a synthetic channel trace. We implemented Proportional Fair Scheduling and Max Weight Scheduling as a $\mu$App in EdgeRIC which exchanges state information and actions with RAN Stack over ZeroMQ-based communication. In order for performance comparison, we implemented Near RT RIC by imposing delays in message deliveries between RIC and RAN. We used synthetic CQI traces for 4 UEs and evaluated each algorithm's performance by comparing their average throughput. The graph on the left side shows average throughput of Proportional Fair Scheduling and the graph on the right side shows one of Max Weight Scheduling. A blue line in the graphs are for EdgeRIC while a red line for Near RT RIC. As Near RT RIC showed low average throughput due to the delay, EdgeRIC successfully supported those microbenchmarks to achieve their best throughput.

\begin{table}[htbp]
\caption{Load: 30Mbps, Channel: 4UE synthetic channel }
    \centering
    \begin{tabular}{|l l c c c|} 
 \hline
  &  & EdgeRIC & 15ms & 30ms \\ 
 \hline
 Max CQI & Avg. Thrpt. & \textbf{26.3} & 16.4&15.9 \\ 
 \hline
  & BL[MB] & \textbf{1.22} & 1.25&1.27 \\
 \hline
 
 Prop. Fair. & Avg. Thrpt. & \textbf{24.27} & 22.72& 20.15 \\ 
 \hline
  & BL[MB] & 1.37 & \textbf{1.08} &1.27 \\
 \hline
 
 Max Weight & Avg. Thrpt. & \textbf{25.4} & 19.8 & 19.1 \\ 
 \hline
  & BL[MB] & 1.33 & \textbf{1.05}&0.99 \\
 \hline
 
\end{tabular}
\label{4UE synthetic}
\end{table}

\begin{table}[htbp]
\caption{Throughput and Backlog Buffer evaluation of synthetic traces}
\centering
\begin{tabular}{l c c c }
\hline
 & PPO & Max Weight & Max CQI \\
\hline
\multicolumn{4}{l}{Synthetic Channel Traces} \\
\hline
Scenario 1& \textbf{27.8/0.38} & 25.6/0.38 & 19.0/0.38 \\
Scenario 2 & \textbf{27.5/0.33}  & 24.7/0.51 & 18.9/0.38\\
Scenario 3& \textbf{26.8/0.38} & 25.2/0.51 & 25.2/0.76 \\
\hline
\end{tabular}
\label{tab:RLevaluation_synth}
\end{table}

\begin{figure}[htbp]
\begin{center}
\includegraphics[width=\columnwidth]{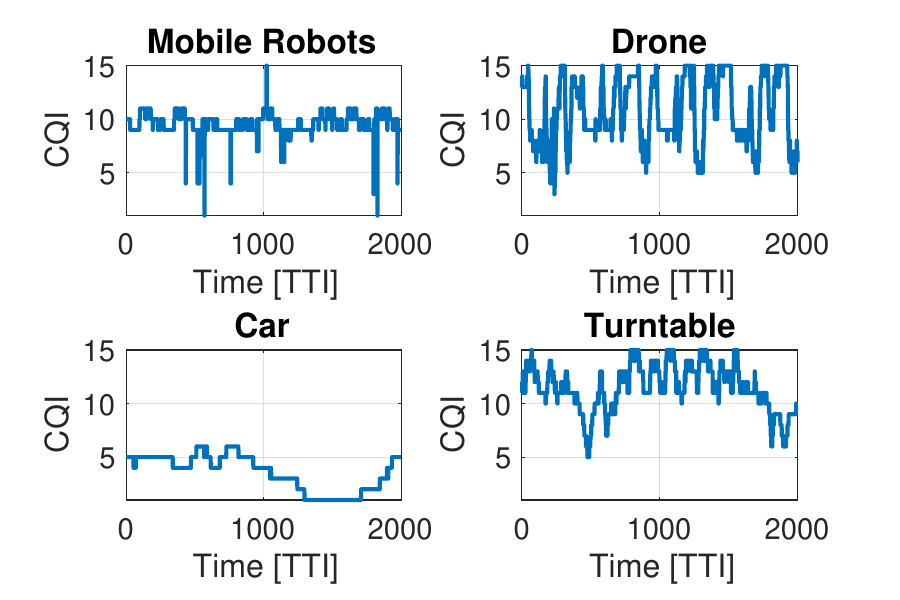}
\caption{CQI Trace for different UEs}
\label{fig:cqi_evol} 
\end{center}
\end{figure}

Figure \ref{fig:cqi_evol} shows the real-time CQI traces we collected to characterize various CQI mobility in different environments. An x-axis is time in TTI unit and a y-axis is CQI showing real-time channel quality. The report period of CQI values was 2 TTI period which is approximately 2 mili-seconds. While a mobile robot has some drops in its CQI traces due to the radio block by the lid of the laptop mounted on the mobile robot when rotating, drone's  swift motions caused radical ups and downs in its CQI traces. In car's CQI traces, its variation has a characteristic of a long period and smooth curves because of its slow acceleration and deceleration. For a turntable, we was able to generate fast angular acceleration which caused drastic and kind of periodic changes in its CQI traces by fast rotation.  


\begin{figure}[htbp]
\begin{center}
\includegraphics[width=\linewidth]{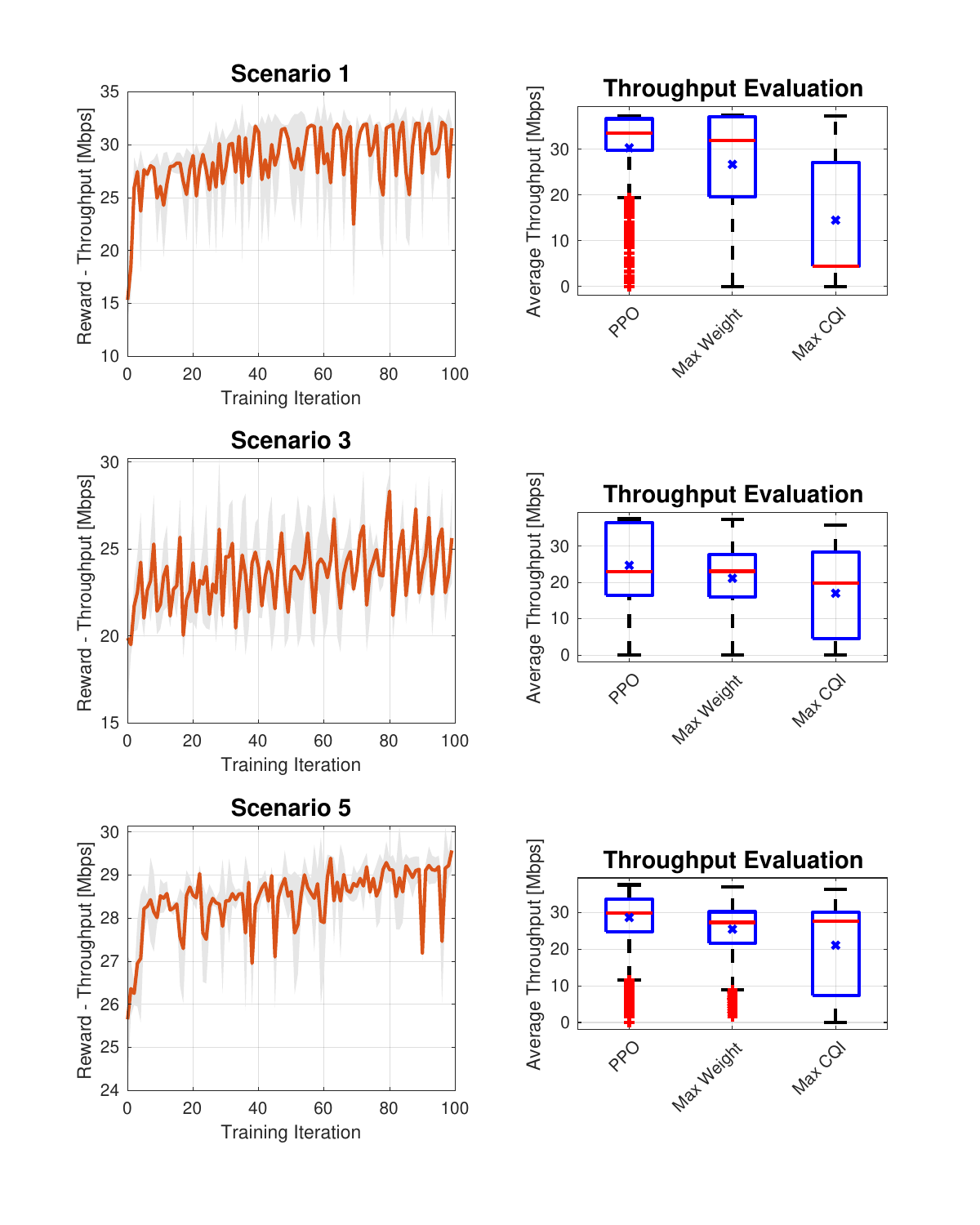}
\caption{Can RL train and evaluate on \twin?}
\label{fig:RLtraining} 
\end{center}
\end{figure}



Figure \ref{fig:RLtraining} shows the training curves of RL PPO policy model and its throughput evaluations on DigitalTwin. The graphs in the first column illustrate the training curve for each scenario. We trained the policy network of RL for Scenario 1, 3 and 5 which are 2 Drone UEs, 2 Car UEs and 2 Drone UEs, and 2 Random Walk UEs and 2 Turntable UEs cases. We used specific CQI trace data to emulate channel conditions for individual scenarios. To train an RL agent using PPO, we designed a reward function to maximize throughput for general scenarios. Each iteration, we sampled 5000 data samples including previous state, previous action and current state at each iteration and updated policy network until the reward saturates. Most transient periods for training were less than 40 iterations and the training curves converged for all scenarios. The graphs in the second column describe the throughput evaluation of the trained policy model. In each graph, we compared the throughput of PPO to the ones of Max Weight and max CQI. In each algorithm, a red bar means the mean of its throughputs and a blue box their range. PPO obviously outperformed Max CQI, and even achieved almost the same throughput performance as Max Weight, which is usually considered as an optimum policy, with somehow higher ranges than Max Weight.



\end{document}